\documentclass[pdflatex,sn-mathphys-ay]{sn-jnl}

\usepackage{setspace}
\usepackage{caption}
\usepackage{graphicx}%
\usepackage{multirow}%
\usepackage{amsmath,amssymb,amsfonts}%
\usepackage{amsthm}%
\usepackage{mathrsfs}%
\usepackage[title]{appendix}%
\usepackage{xcolor}%
\usepackage{textcomp}%
\usepackage{manyfoot}%
\usepackage{booktabs}%
\usepackage{algorithm}%
\usepackage{algpseudocode}%
\usepackage{listings}%
\usepackage{subcaption}
\usepackage{latexsym}%
\usepackage{siunitx}%
\usepackage{hyperref}%
\usepackage{threeparttable}%
\usepackage{adjustbox}

\usepackage{mathtools}%
\mathtoolsset{showonlyrefs=true}

\theoremstyle{thmstyleone}%
\theoremstyle{thmstyletwo}%

\theoremstyle{thmstylethree}%

\begin{document}

\title[]{Integrating Deep Learning and Spatial Statistics in Marine Ecosystem Monitoring}

\author*[1]{\fnm{Gian Mario} \sur{Sangiovanni}}

\author[2]{\fnm{Gianluca} \sur{Mastrantonio}}

\author[3]{\fnm{Alessio} \sur{Pollice}}

\author[4]{\fnm{Daniele} \sur{Ventura}}

\author[1]{\fnm{Giovanna} \sur{Jona Lasinio}}

\abstract{In ecology, photogrammetry is a crucial method for efficiently collecting non-destructive samples of natural environments. When estimating the spatial distribution of animals, detecting objects in large-scale images becomes crucial. Object detection models enable large-scale analysis but introduce uncertainty because detection probability depends on various factors. To address detection bias, we model the distribution of a species of benthic animals (holothurians) in an area of the Italian Tyrrhenian coast near Giglio Island using a Thinned Log-Gaussian Cox Process (LGCP). We assume that a "true" intensity function accurately describes the distribution, while the observed process, resulting from independent thinning, is represented by a 	\textit{degraded} intensity. The detection function controls the thinning mechanism, influenced by the object's location and other detection-related features.
We use manual identification of holothurians as our benchmark. We compare automatic detection with this benchmark, an unthinned LGCP, and the thinned model to highlight the improvements gained from the proposed approach.Our method allows researchers to use photogrammetry, automatically identify objects of interest, and correct biases and approximations caused by the observation process.
}

\keywords{Thinned Log-Gaussian Cox Process, Detection Bias, Underreporting, Object Detection, Marine Biology}



\maketitle

\section{Introduction}\label{sec1}

Modern underwater imaging technology has created both an opportunity and a challenge for marine ecology. Advances in Structure from Motion (SfM) photogrammetry, when paired with Diver Propulsion Vehicles (DPVs), now enable researchers to capture high-resolution imagery across hundreds of square meters of seafloor in a single survey \citep{ventura2025detecting}. A typical large-scale benthic survey can generate thousands of images (once tiled in a suitable format), each containing individual organisms. This represents an unprecedented opportunity to study marine populations at spatial scales that were simply unattainable with traditional survey methods. However, this technological capability has revealed that our analytical capacity has not kept pace with our data collection capabilities. The standard approach to analysing underwater imagery remains manual annotation by trained biologists, who visually inspect each image and mark the location and identity of target organisms. While this method provides reliable, high-quality data, it is fundamentally limited by its labour-intensive nature \citep{beijbom2012automated, mahmood2016automatic}. A single identification process on a large-scale image could require several weeks of continuous work to analyse completely.  We can now survey vast areas of the seafloor in hours, yet we lack the analytical capacity to process the resulting imagery within a reasonable timeframe. The bottleneck has shifted from data collection to data analysis, effectively preventing us from monitoring benthic ecosystems at the spatial and temporal scales necessary to understand ecological dynamics and support conservation planning.

Deep learning appears to offer a solution. Convolutional neural networks (CNNs) have demonstrated impressive capabilities for automated species detection and counting in ecological imagery \citep{weinstein2018computer, de2022special}. These methods can process thousands of images in minutes and often achieve detection accuracy comparable to human annotators on benchmark datasets. The workflow seems straightforward. Train a detector on manually annotated images, then deploy it across entire surveys to generate organism counts at scales that would be impossible with manual annotation alone. Yet most applications of deep learning in ecology treat detection as the final product rather than as the beginning of ecological inference. Researchers focus primarily on maximising detection accuracy (or related metrics). While these performance measures matter for evaluating detectors, they sidestep two fundamental issues that determine whether automated detection can actually support robust ecological conclusions. First of all, all automated detection systems imperfectly observe ecological reality. Even the most sophisticated detectors make systematic mistakes. They miss cryptic or partially obscured individuals, perform poorly under challenging lighting conditions or in turbid water, confuse target species with morphologically similar organisms or background structures, and show inconsistent performance across different habitats or image qualities \citep{zhai2022underwater}. These errors are not random noise that cancels out over large samples. They represent systematic observation bias. When we treat detector outputs as ground truth for estimating abundance, these biases translate directly into biased ecological conclusions. Secondly, beyond the detection accuracy problem lies a deeper conceptual issue about what we actually want to learn from imagery data. The spatial distribution of organisms reflects underlying ecological processes. Benthic species distribute themselves across seascapes according to environmental gradients in depth, slope, and substrate type, responding to habitat heterogeneity, interacting with other species, and reflecting historical colonisation dynamics. These spatial patterns constitute the primary scientific interest in most ecological studies. They reveal which habitats species prefer, identify the environmental factors that structure populations, and enable predictions about how distributions might shift in response to environmental change. Most detection-focused approaches collapse this rich spatial information into simple aggregate counts. The patterns that ecologists most want to understand get discarded in the process of counting objects. For those reasons, we need a different approach, one that treats automated detection not as ground truth but as an observation process that must be modelled explicitly within a larger inferential framework. This perspective has long been central to ecological statistics. The distinction between actual ecological state and observed data forms the conceptual foundation for species distribution models, capture-recapture analyses, and occupancy models \citep{mackenzie2002estimating,warton2013model}. All these methods explicitly separate the ecological process (where organisms actually occur and how many are present) from the observation process (what we manage to detect given imperfect sampling). The same logic should be applied in this context.

We developed this approach here for sea cucumber populations assessed through large-scale underwater photographic surveys. Sea cucumbers (Holothuroidea) make an excellent model system for several reasons. Ecologically, they play a key ecological role in benthic dynamics where they are involved as
ecosystem engineers in the processing of organic matter in the detrital food web pathway \citep{ciriminna2024aquaculture}. They process large quantities of sediment and contribute substantially to nutrient cycling and bioturbation in benthic communities ranging from shallow seagrass beds to deep-sea environments \citep{purcell2016ecological, schneider2013inorganic, lopez1987ecology}. Many sea cucumber populations face conservation concerns due to intensive fishing pressure driven by commercial markets (predominantly Asian). From a methodological standpoint, sea cucumbers present typical challenges for automated detection. They show variable appearance, exhibit cryptic colouration, resemble background substrate and occur in visually complex habitats. Methods that work for sea cucumbers should generalise reasonably well to other benthic species facing similar detection challenges.

Our approach divides the problem into two stages that we can address with appropriate methods while carefully modelling how they connect. The first stage uses a deep learning object detector trained on a manually annotated set of sea cucumber images. This detector functions as an automated annotator that processes large volumes of imagery to identify presence locations. We do not treat these detections as error-free observations of where sea cucumbers actually occur. Instead, we recognise them as a thinned sample of the actual underlying population, a partial and biased representation shaped by the detector's imperfect observation capabilities. The second stage models the spatial distribution of detected organisms using a thinned Log-Gaussian Cox Process (LGCP) \citep{moller1998log, chakraborty2011point, warton2010poisson}. This formulation naturally captures complex spatial patterns and their relationships with environmental covariates. Our key innovation involves explicitly incorporating a thinning mechanism. We model the detections as arising from an underlying true point pattern (actual sea cucumber locations) that has been stochastically thinned. The thinning probability reflects detection reliability and can vary spatially, depending on factors such as image quality, habitat complexity, and others.
This thinning framework provides the natural connection between our two stages. The detector provides us with presence-only data, without indicating the detection probability. The spatial model estimates this detection probability while simultaneously recovering the actual underlying spatial intensity and quantifying its relationship with environmental covariates. We can therefore correct for underdetection and obtain unbiased estimates of total abundance and spatial distribution patterns. The framework also provides appropriate uncertainty quantification throughout.

This approach inverts the relationship between detection and spatial modelling found in existing computer vision research. Several recent studies have embedded spatial point process models directly into detection networks \citep{pham2016efficient, mabon2022point, mabon2023learning, mabon2024learning, descombes2002marked}. Due to their flexibility and scalability, neural networks and spatial point process models have been increasingly recognised as a \textit{convenient couple} for the statistical analysis of multivariate point patterns \citep{mateu2022spatial}. These methods use spatial statistics as regularisation tools. They incorporate prior knowledge about expected spatial patterns to improve detection accuracy on individual images or image sequences. The spatial model serves the detector, and the goal is to maximise accuracy.

\begin{figure}[h!]
\centering
  \includegraphics[width = 0.8\textwidth, height = 0.7\textheight,keepaspectratio]{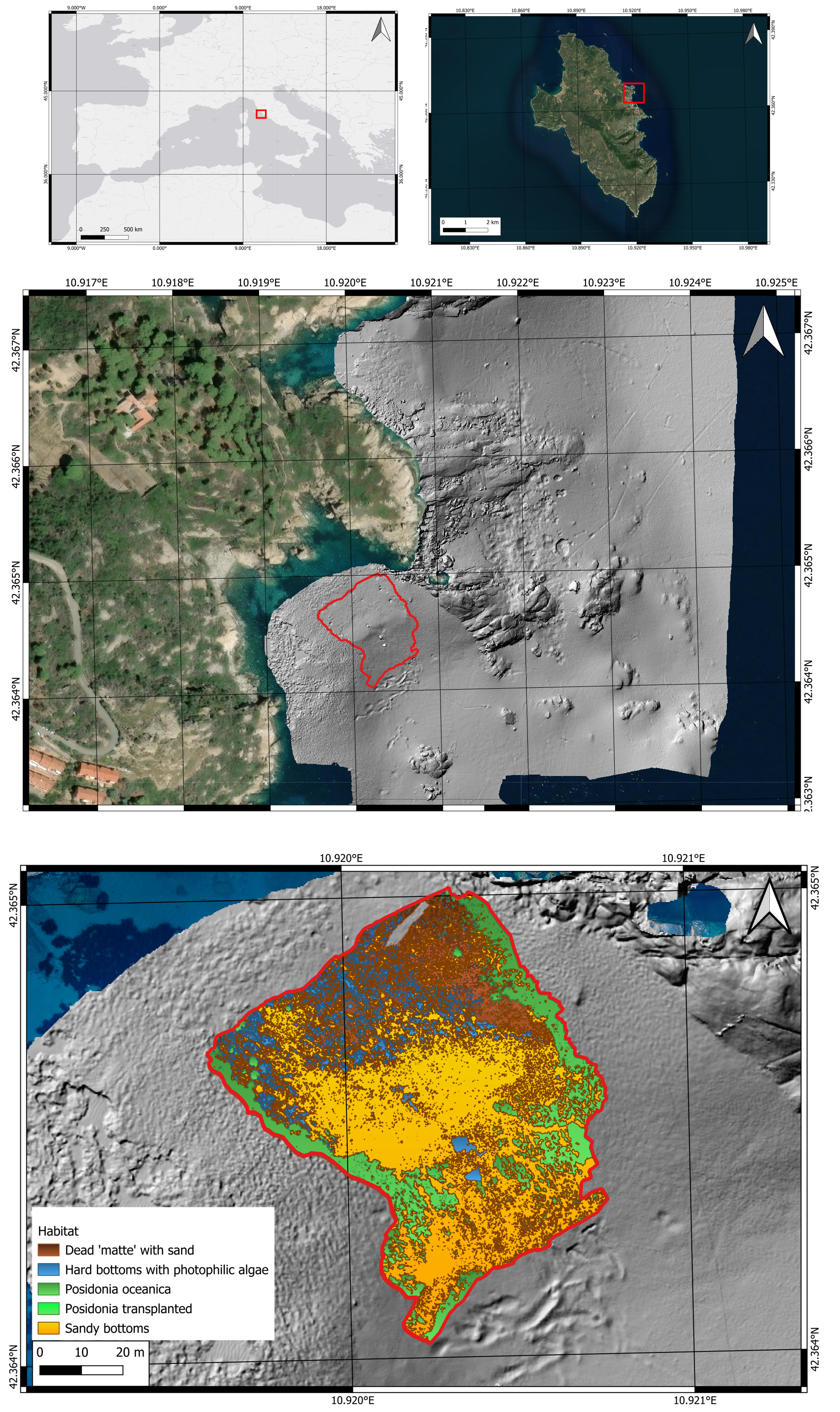}%
  \caption{Punta Gabbianara study site.}\label{Fig_Extent}
\end{figure}

Separating detection from inference in this modular way offers several practical benefits. The approach scales naturally to large spatial extents without retraining detectors for each new study area or survey period. A single trained detector can be applied across multiple sites and time periods, with the spatial model adapting to local environmental conditions through covariate effects. The framework allows direct incorporation of environmental data to model the ecological drivers of spatial patterns. Comprehensive uncertainty quantification becomes possible, propagating uncertainty from both the detection process and the spatial process. Perhaps most importantly, the framework enables genuine ecological inference rather than treating detection as an endpoint.

The remainder of this paper proceeds as follows. \hyperref[sec2]{Section~\ref*{sec2}} describes the study area and details both stages of our framework: the object detection phase and the thinned LGCP model, including environmental covariate preprocessing and model selection procedures. \hyperref[sec3]{Section~\ref*{sec3}} presents results comparing alternative model specifications and demonstrates how our approach recovers spatial patterns and abundance estimates from detector outputs. \hyperref[sec4]{Section~\ref*{sec4}} discusses the ecological implications, methodological limitations, and potential extensions of this framework to other benthic monitoring applications.

\section{Materials and methods}\label{sec2}

\subsection{Study Area}

The research was conducted at Punta Gabbianara (N: 42.364867\textdegree; E: 10.920210\textdegree), situated along the northeastern shoreline of Giglio Island in the Tyrrhenian Sea, Italy (Figure \ref{Fig_Extent}). The study area encompasses approximately $5.500$ square meters of infralittoral seafloor, extending from $8$ to $27$ meters depth, and is characterised by heterogeneous benthic habitats that have experienced substantial anthropogenic impacts. This location became the focus of intensive scientific interest following the Costa Concordia maritime disaster in 2012 \citep{casoli2017assessment}. The grounded vessel and subsequent recovery operations created extensive shading effects that severely impacted the endemic \textit{Posidonia oceanica (L.) Delille} (P. Oceanica) seagrass meadows, leading to substantial habitat degradation and reduced vegetative cover \citep{mancini2019impact, toniolo2018seagrass}. Following the wreck's removal in July $2014$ and the completion of environmental remediation activities in April $2018$, an ambitious five-year ecological restoration initiative commenced in $2019$, focusing on large-scale \textit{P. oceanica} transplantation and habitat reconstruction \citep{mancini2022transplantation}.

The current benthic landscape represents a complex mosaic of Mediterranean infralittoral communities, characterised by sandy sedimentary areas interspersed with granitic geological formations, including boulders and cobbles colonised by photophilic macroalgal assemblages. The site exhibits both remnant natural \textit{P. oceanica} patches and newly established transplanted meadows, which are progressively colonising fragmented areas of dead rhizome matrices (locally termed ``matte'') and adjacent sandy substrates. This heterogeneous habitat configuration provides an ideal setting for investigating benthic invertebrate distributions across varying substrate types and restoration stages, making it particularly suitable for automated detection and spatial modelling approaches.

 \subsection{Data Collection}

\begin{figure}[t]
    \centering
     \subfloat[]{\includegraphics[scale = 0.36]{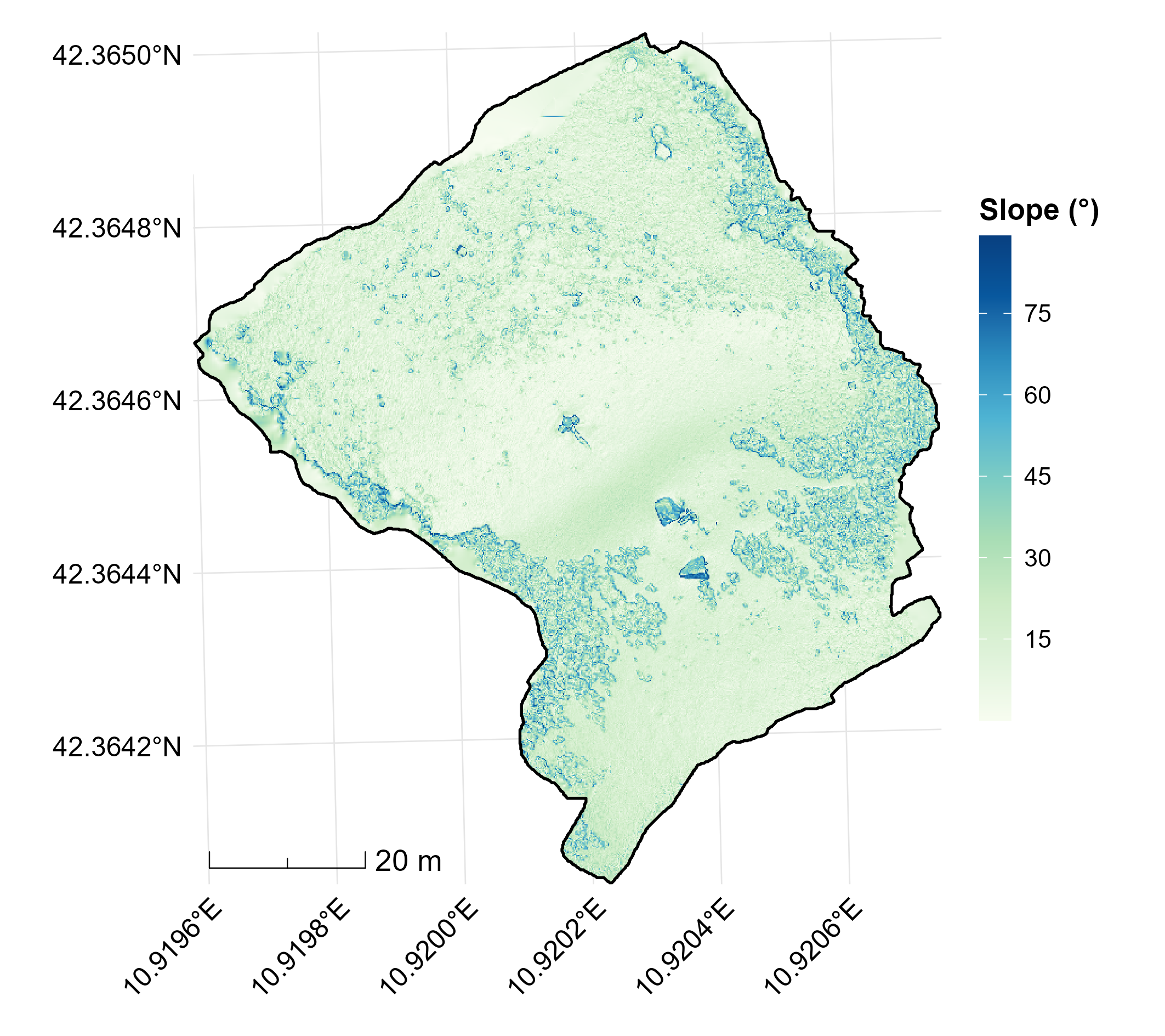}}
     \subfloat[]{\includegraphics[scale = 0.36]{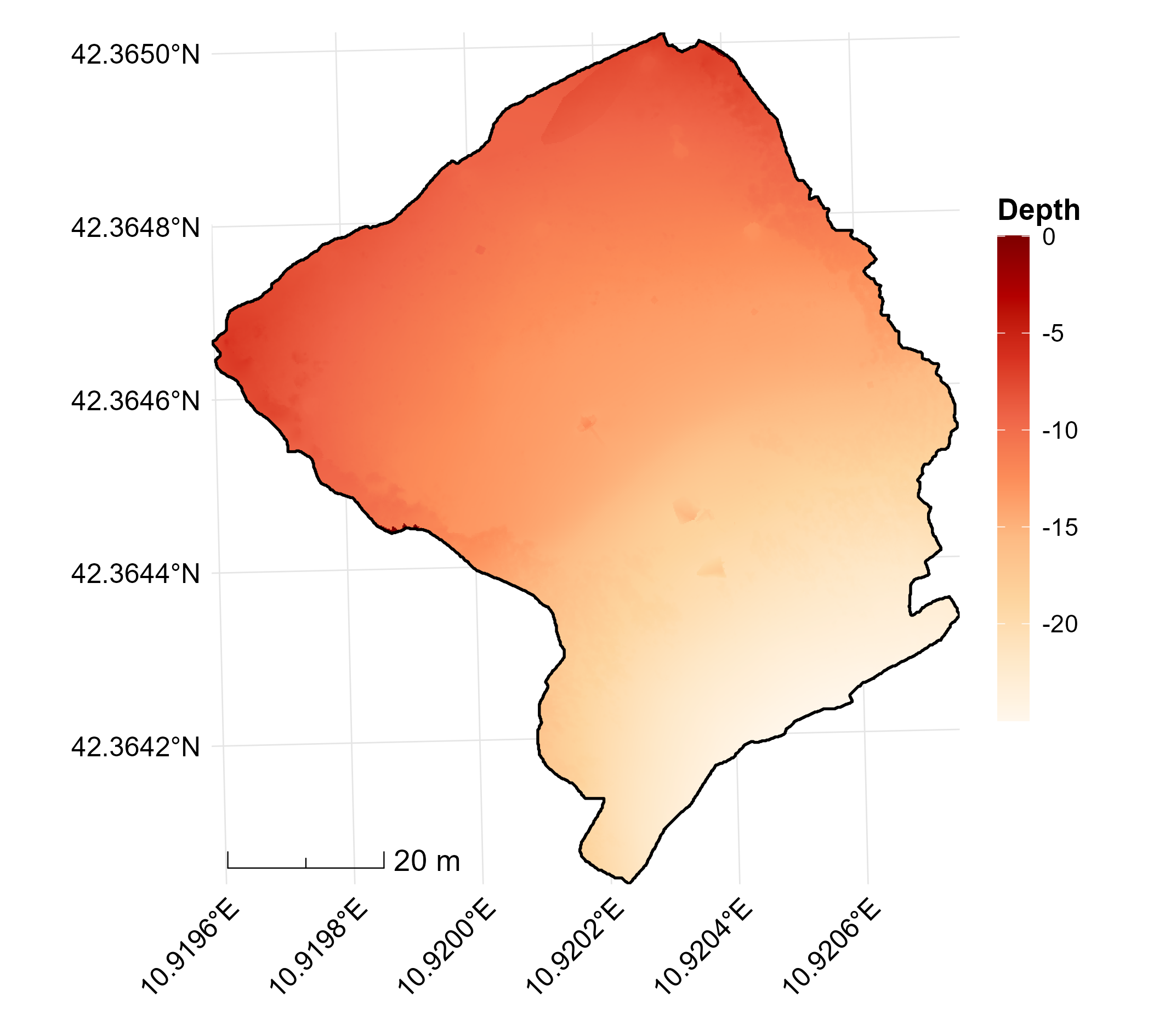}}\\
          \subfloat[]{\includegraphics[scale = 0.3]{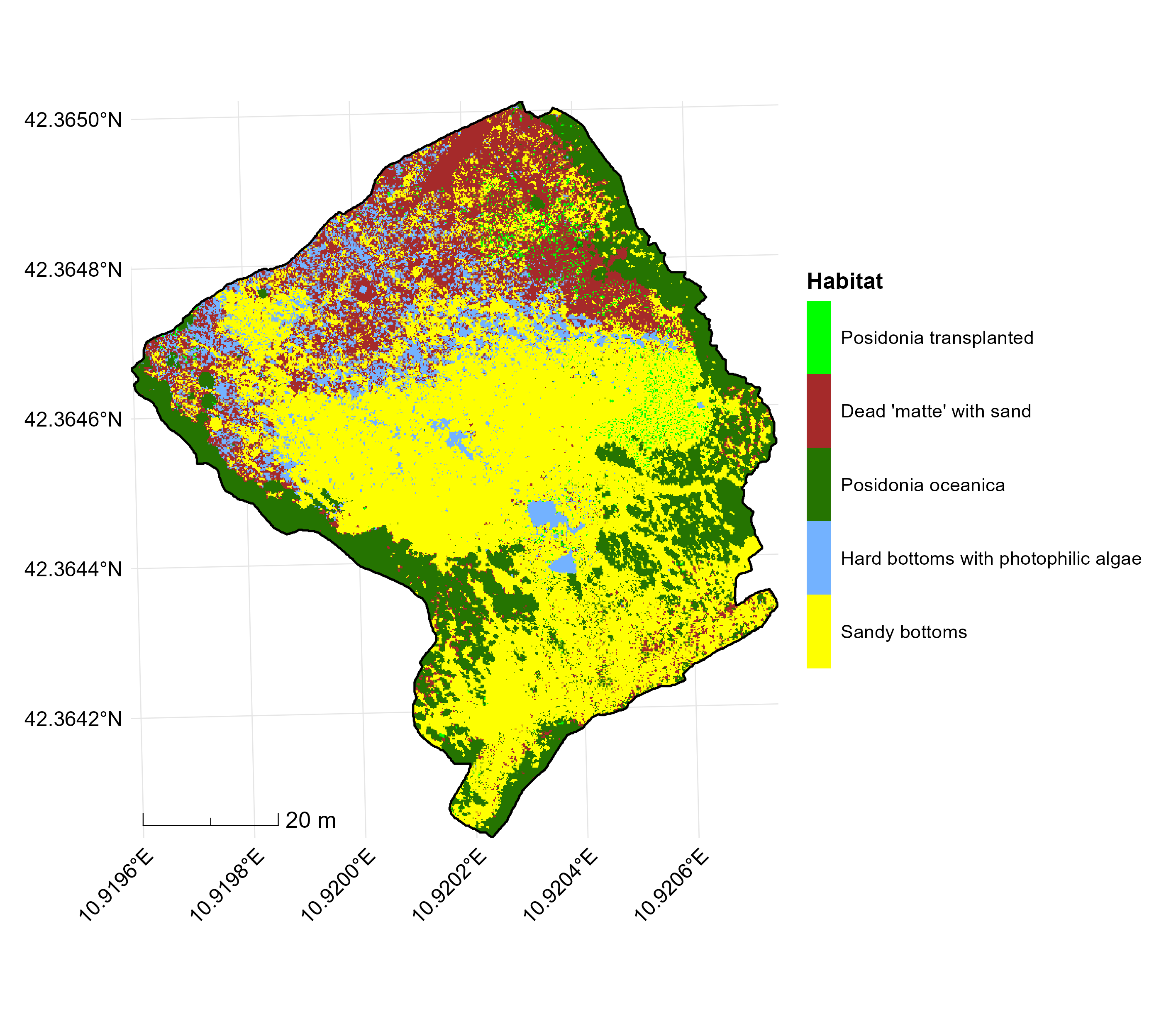}}
     \subfloat[]{\includegraphics[scale = 0.25]{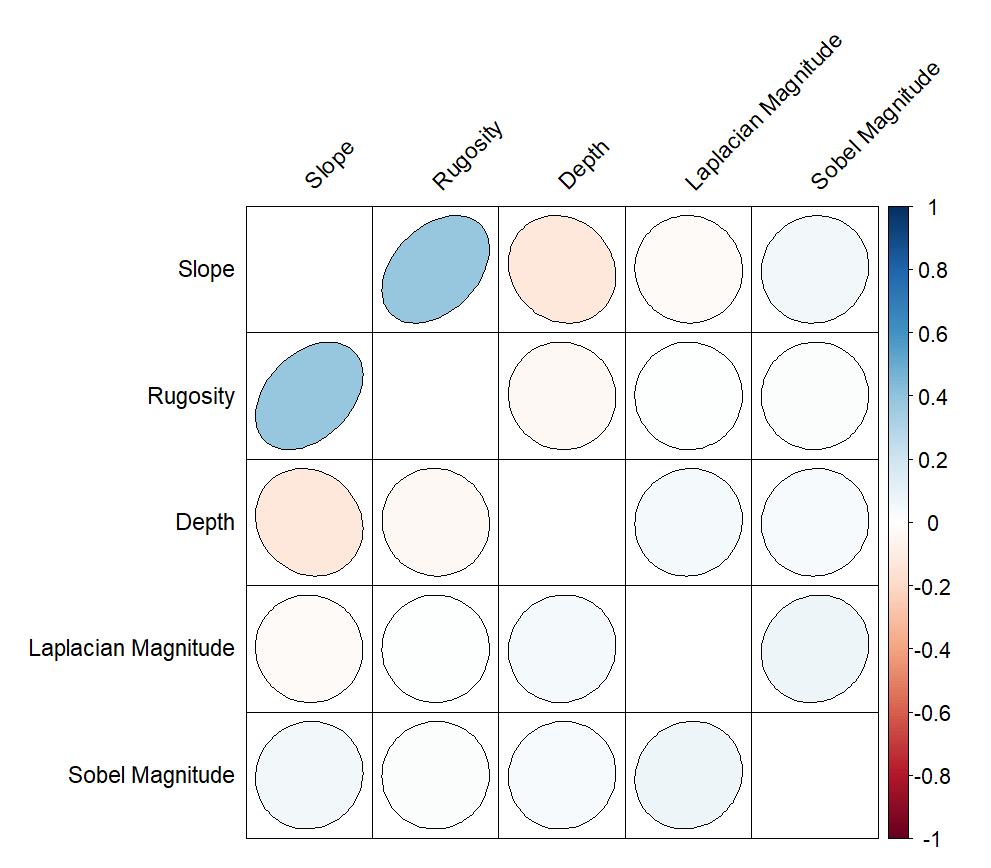}}
     \caption{Spatial distribution of (a) slope, (b) depth, and (c) habitat types within the study area. Panel (d) shows the correlation matrix among environmental covariates.}
     \label{fig:eda1}
\end{figure}

During $2022$, we conducted five photogrammetric mapping missions spanning January through November, applying methodologies previously validated for tracking seagrass transplant distribution \citep{ventura2022high} and characterising sea cucumber microhabitat selection \citep{ventura2025detecting}.  Underwater image acquisition employed a GoPro Hero 10 action camera ($23$ MP,$5568 \times 4176$ pixel resolution) affixed to a DPV, enabling extensive spatial coverage while constraining dive duration. Camera positioning at roughly $5$ meters above the seabed maintained $75\%$ image overlap between successive frames, yielding $3$ mm pixel ground sample distance which was sufficient for organism detection and substrate characterisation.

We incorporated the same suite of spatial  variables utilised in previous analyses \citep{mastrantonio2024species, ventura2025detecting}, maintaining a resolution of $0.21 \times 0.24 \ m$. Continuous environmental variables, including seafloor depth, slope angle and terrain roughness, which were all derived from the high-resolution digital surface models. Given the strong correlation between terrain roughness and slope, we retained only slope in subsequent analyses (Figure \ref{fig:eda1}). Beyond the variables used in prior studies, we enhanced our feature set by applying edge detection filters to the orthophotomosaics. Specifically, we applied Laplacian and Sobel operators \citep{robinson1977edge} to capture complementary aspects of seafloor texture. The Laplacian (second-derivative filter) highlights fine-scale textural discontinuities and blob-like features, while the Sobel operator (first-derivative filter) emphasises directional edges and linear boundaries. For the Sobel filter, we computed horizontal and vertical gradients separately, then calculated edge magnitude as their dot product. Both Laplacian and Sobel magnitude values were averaged across Red-green-blue channels to produce single-band texture metrics. These textural variables may capture ecologically relevant habitat features such as rubble-sand interfaces or algal canopy boundaries that are not adequately represented by topographic metrics alone. 

To classify substrate types from the exceptionally detailed orthophoto mosaics, we applied an object-based image analysis framework \citep{ventura2025detecting,fallati2024combining}. This approach first segments imagery into discrete units sharing similar spectral signatures and spatial characteristics, then applies supervised machine learning to assign each unit to a habitat category. Five benthic classes were identified: sandy bottoms, hard bottoms with photophilic algae, dead matte mixed with sand, natural \textit{P. oceanica} meadow, and transplanted \textit{P. oceanica}. All environmental variables were aggregated to a regular $1 \times 1$ m grid using zonal averaging. For each grid cell, we derived two representations of habitat information: the fractional coverage of each habitat type and the dominant habitat class. This dual representation enables subsequent models to capture both fine-scale habitat heterogeneity and the primary substrate composition at the analysis resolution.


\subsection{Detection Task}

To automate the identification of sea cucumbers within orthophotomosaics, we developed an object detection workflow based on the You Only Look Once version~11 (YOLOv11) architecture \citep{redmon2016you}. This single-stage detector was selected for its optimal balance of computational efficiency and high accuracy, which is crucial for processing extensive seabed imagery while reliably identifying small objects \citep{unel2019power}. The YOLOv11 architecture comprises a backbone for feature extraction, a neck for feature aggregation, and a head for final predictions, optimised via a composite loss function.


The training dataset was derived from twelve high-resolution orthophotomosaics acquired in~$2021$ at ecologically similar sites outside the main study area, including both transplantation sites and natural habitats such as \textit{Cala Cupa} and \textit{Cala di Mezzo} along the northeastern coast of the island. To match the scale of the target organisms, each mosaic was tiled into $640 \times 640$~pixel patches at a spatial resolution of $0.5$~cm/pixel using the QGIS plugin Deepness \citep{aszkowski2023deepness}. From an initial pool of $3609$ tiles, we curated a dataset of $2492$ annotated images, including $2292$ positive tiles and $200$ negative samples depicting only the seabed. The inclusion of these negative examples, spanning various habitats, was a strategic choice to enhance the model's generalisation capacity and reduce false positives. A rigorous, dual-reviewer annotation process on the Roboflow platform \citep{alexandrova2015roboflow} ensured high-quality labels. The dataset was partitioned into training ($70\%$), validation ($20\%$), and test ($10\%$) sets.

\begin{figure}[t]
    \centering
     \subfloat[]{\includegraphics[scale = 0.27]{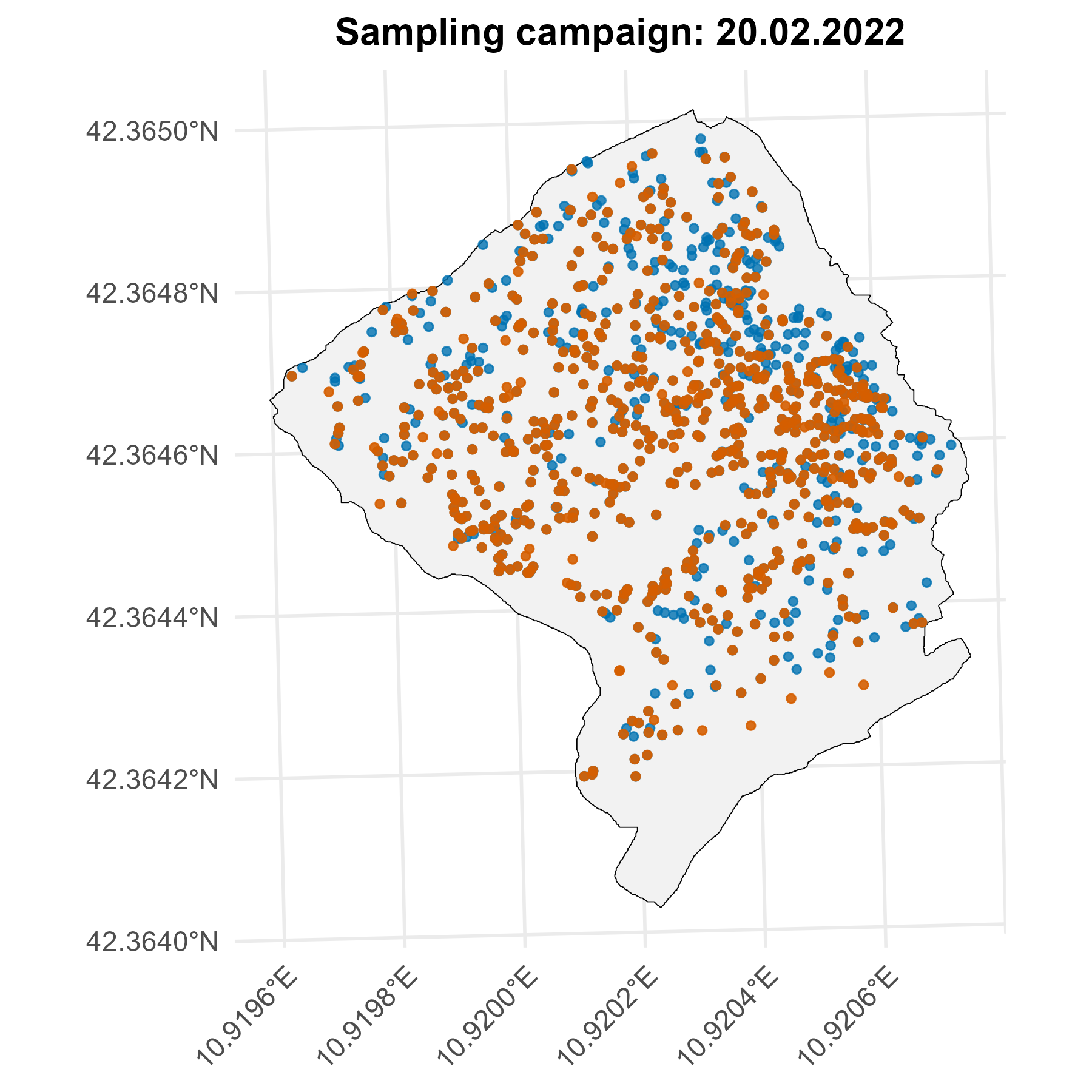}}
     \subfloat[]{\includegraphics[scale = 0.27]{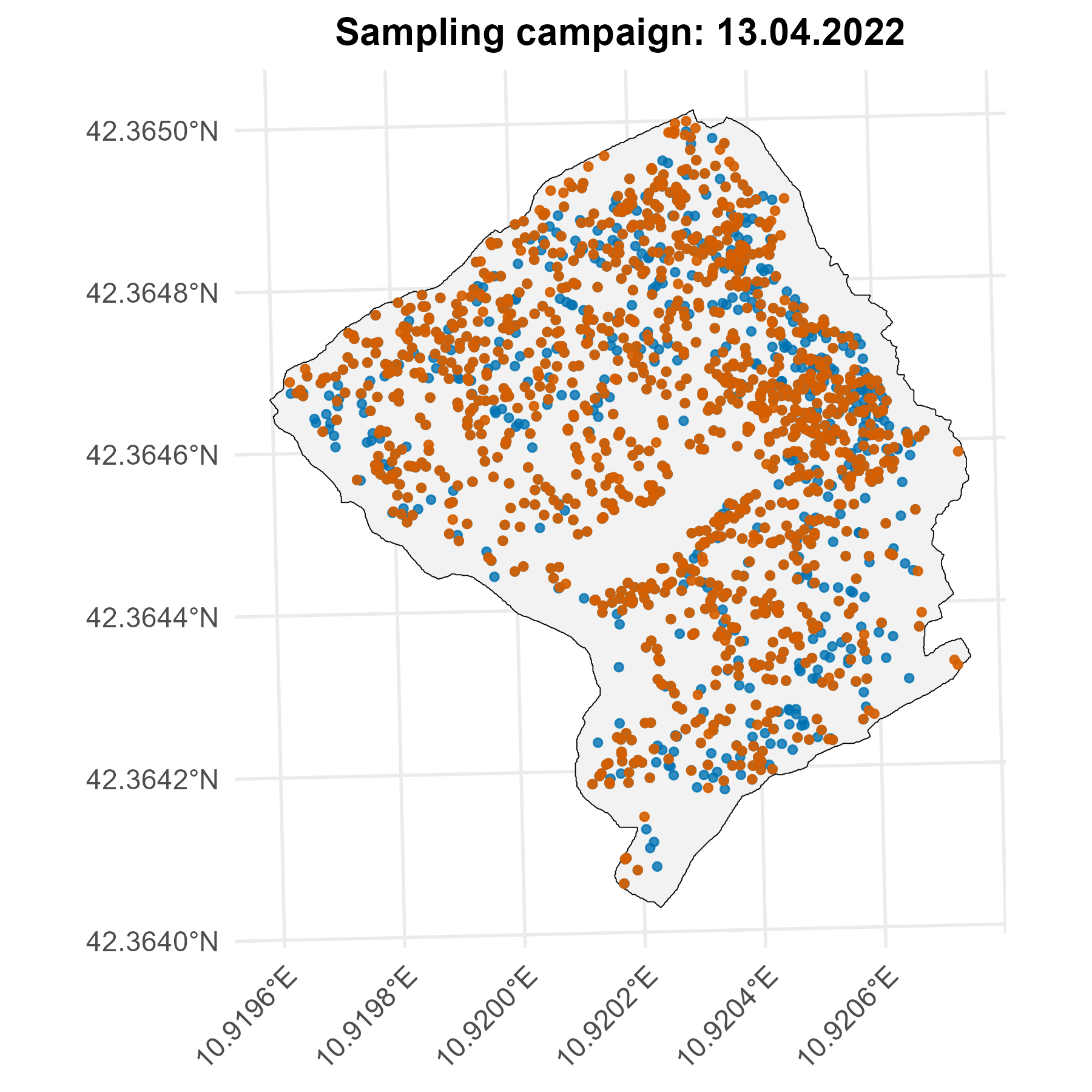}}
     \subfloat[]{\includegraphics[scale = 0.29]{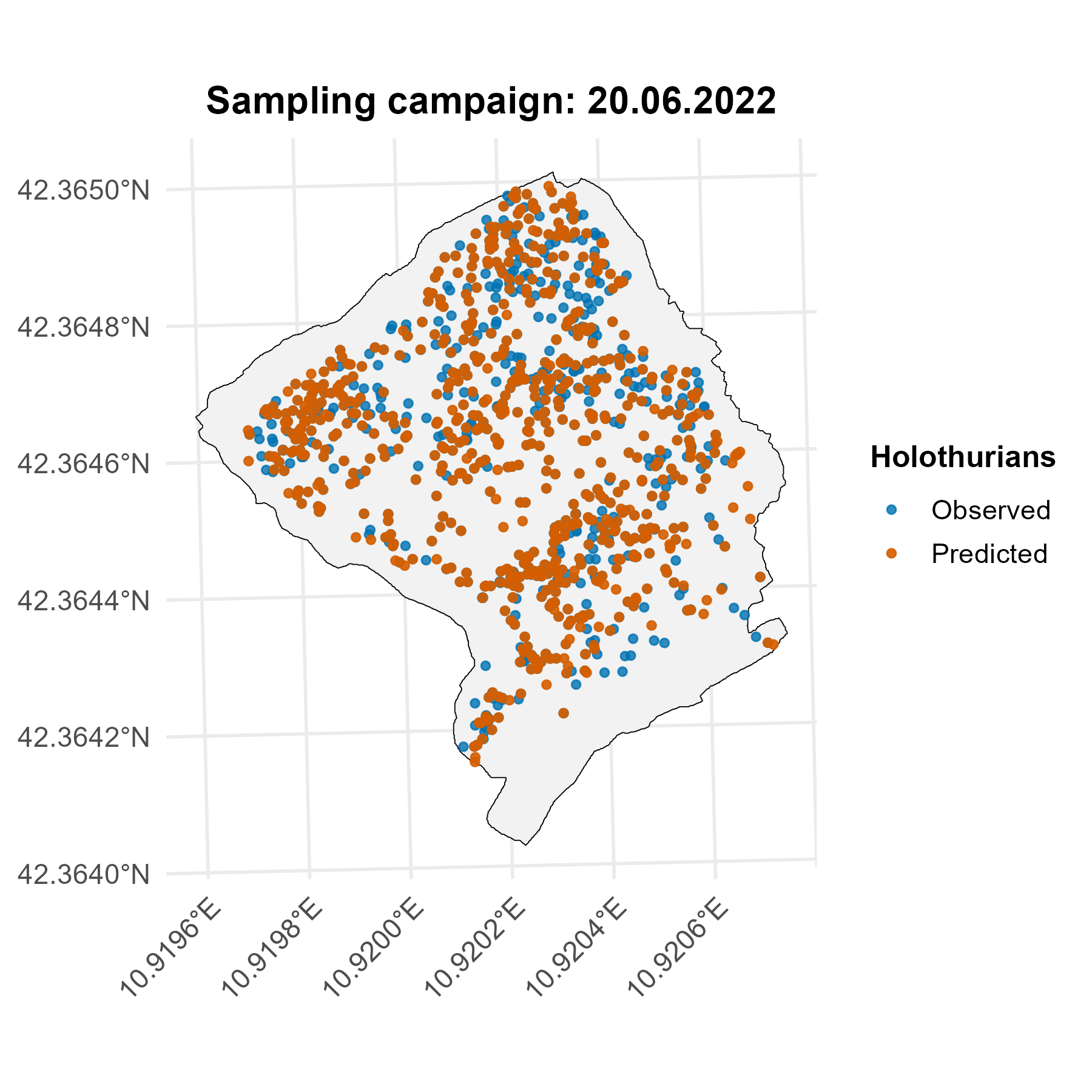}}\\
     \subfloat[]{\includegraphics[scale = 0.27]{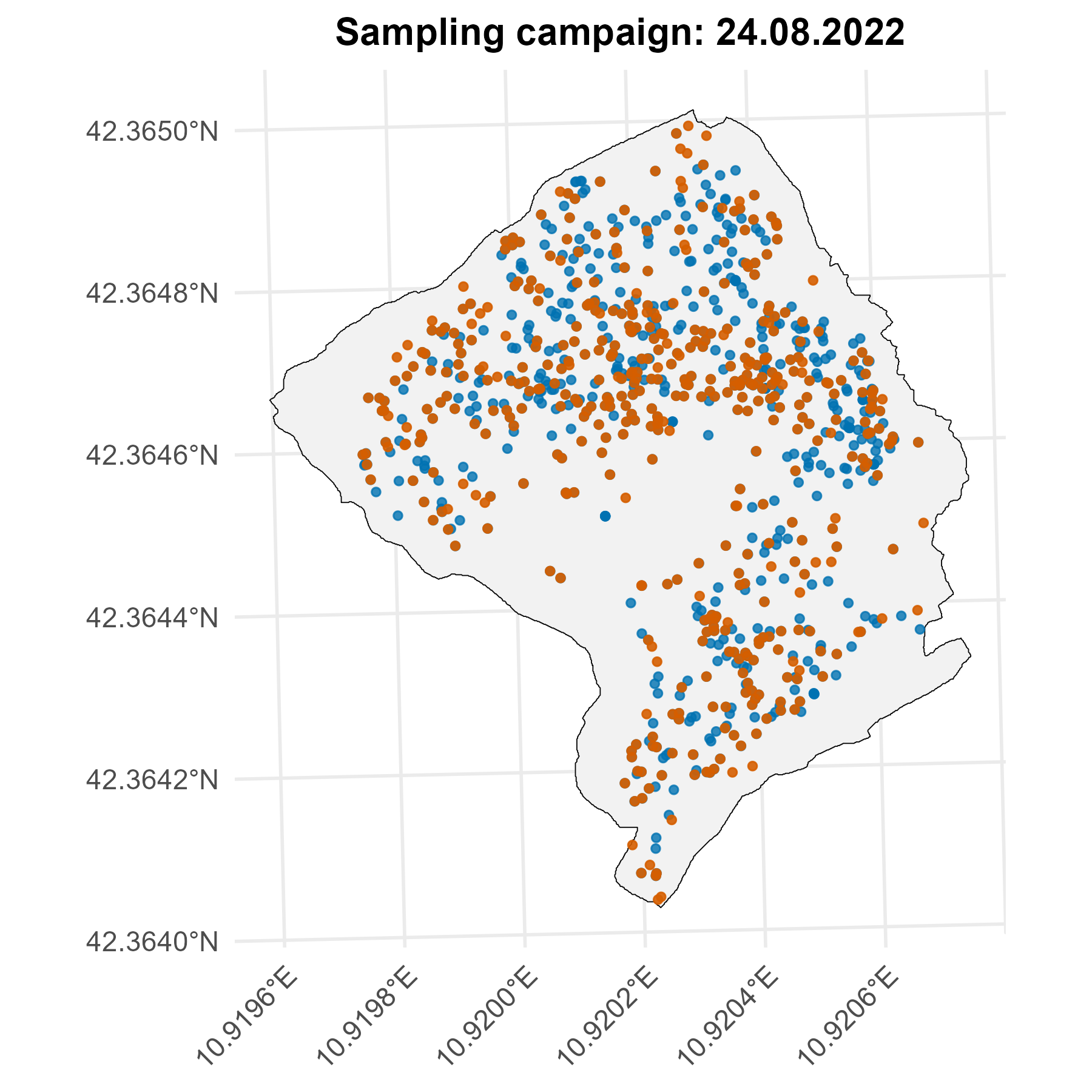}}
     \subfloat[]{\includegraphics[scale = 0.29]{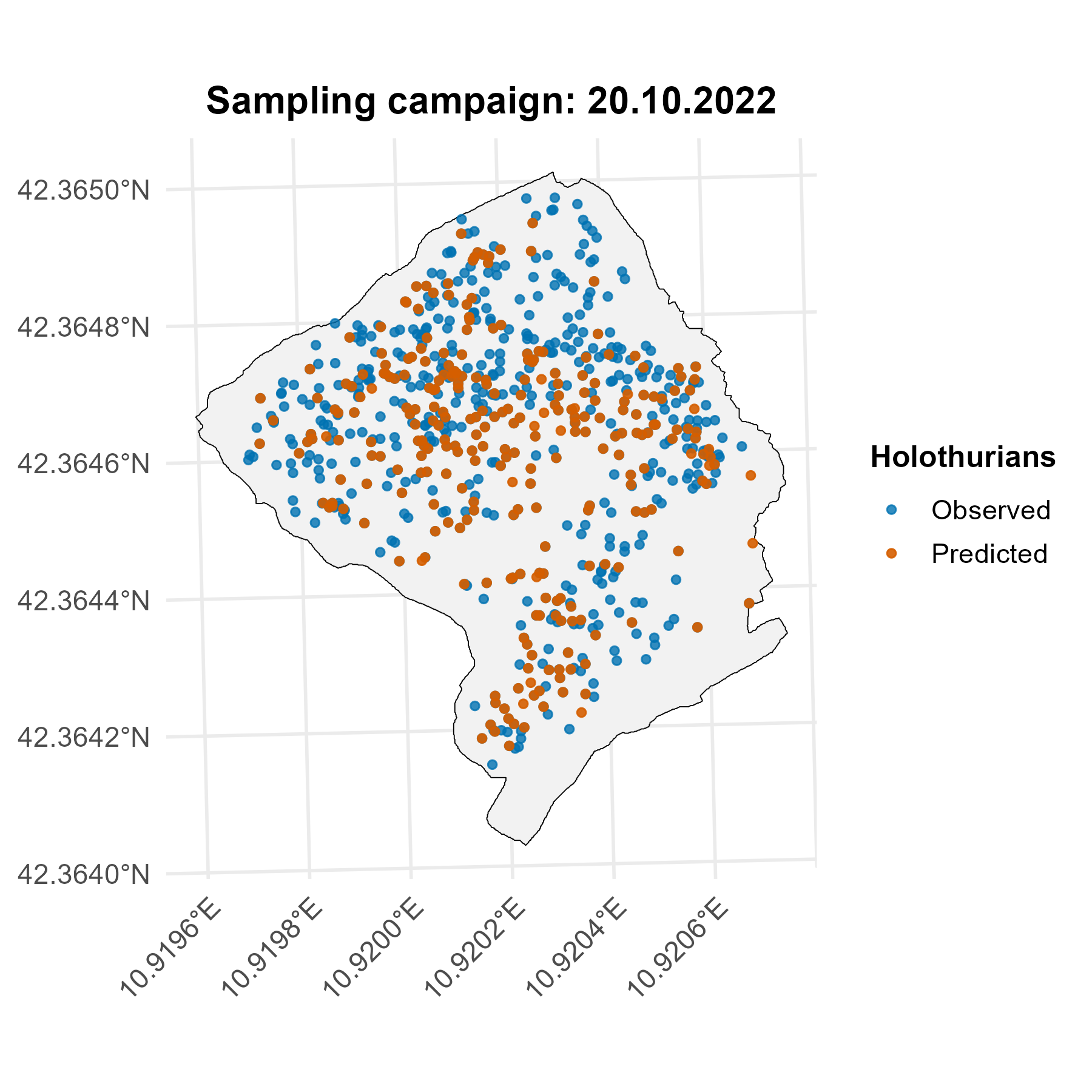}}
     \caption{Distribution of Observed and detected points by the object detector in the study area for the five different campaigns.}
     \label{fig:habitat}
\end{figure}



To further enhance robustness and generalisation capabilities, a comprehensive suite of data augmentation techniques was applied. These included geometric transformations (e.g. rotations, flips, scaling and cropping) and photometric adjustments (e.g. variations in brightness, contrast, hue, and saturation) designed to simulate realistic changes in underwater imaging conditions. Additional effects such as blur and noise were incorporated to reproduce the impact of water turbidity and camera motion. This combination of transformations aimed to improve the model’s resilience to variations in seabed texture, lighting conditions and organism orientation \citep{zoph2020learning}. For the detection task, we employed the medium-sized YOLOv11 model (YOLOv11m), which offers a favourable trade-off between performance and speed. The model was initialised via transfer learning using weights pre-trained on the  Common Objects in Context (COCO) dataset \citep{lin2014microsoft}, a large-scale benchmark comprising natural images of $80$ object categories that facilitates feature generalisation across visual domains. The confidence threshold for predictions was selected by maximising the $F_{1}$ score on the validation set. Further methodological details, including training procedures, hyperparameter optimisation, and test performance, are reported in \citep{Sangiovanni2025}.




The trained model was applied to the five orthophoto campaigns from the main study area. Figure~\ref{fig:habitat} illustrates the spatial distribution of observed versus detected individuals, while Table~\ref{tab:detection_performance} summarises the results. Performance was strong across most campaigns, though a notable decline in recall was observed during the autumn campaign. This decline was primarily attributed to the heightened accumulation of debris, including dead Posidonia leaves, which were scattered across the seabed following storms
that occurred at the end of summer. This situation resulted in the challenging detection of
sea cucumbers.


\begin{table}[t]
\caption{Detection performance across five monitoring campaigns. Habitat-specific counts show observed and predicted individuals. Performance metrics include precision, recall, and $F_1$-score.}
\label{tab:detection_performance}
\centering
\footnotesize
\setlength{\tabcolsep}{1pt} 
\renewcommand{\arraystretch}{1.3}
\begin{tabular}{lccccccccc}
\toprule
& \multicolumn{5}{c}{\textbf{Observed / Predicted counts by habitat}} & \multicolumn{4}{c}{\textbf{Performance metrics}} \\
\cmidrule(lr){2-6} \cmidrule(lr){7-10}
\textbf{Campaign} & \textbf{P.~Transpl.} & \textbf{Dead Matte} & \textbf{P.~Oceanica} & \textbf{Hard} & \textbf{Sandy} & \textbf{Manual} & \textbf{Auto} & \textbf{P} & \textbf{R / F1} \\
\midrule
20/02/2022 & 35 / 23 & 189 / 108 & 54 / 35 & 90 / 56 & 616 / 488 & 985 & 721 & 0.900 & 0.659 / 0.761 \\
13/04/2022 & 302 / 174 & 282 / 210 & 250 / 157 & 433 / 299 & 235 / 301 & 1479 & 1137 & 0.916 & 0.704 / 0.796 \\
20/06/2022 & 20 / 8 & 227 / 163 & 39 / 30 & 101 / 55 & 562 / 440 & 938 & 672 & 0.880 & 0.638 / 0.743 \\
24/08/2022 & 30 / 9 & 171 / 81 & 13 / 10 & 103 / 48 & 468 / 328 & 770 & 472 & 0.826 & 0.506 / 0.628 \\
20/10/2022 & 15 / 2 & 151 / 52 & 21 / 10 & 96 / 23 & 388 / 197 & 668 & 284 & 0.908 & 0.386 / 0.542 \\
\bottomrule
\end{tabular}
\end{table}

\subsection{Data Modelling}

After applying the object detection algorithm to each of the five monitoring campaigns ($t = 1, \ldots, 5$), we obtained two sets of spatial locations for each campaign: (i) manually annotated ground truth positions $\mathbf{U}_t = \{\mathbf{u}_{1t}, \ldots, \mathbf{u}_{m_{t}t}\} \subset \mathcal{D}$, where $m_{t}$ denotes the number of manually identified individuals, and (ii) automated detections $\mathbf{U}^*_t = \{\mathbf{u}^*_{1t}, \ldots, \mathbf{u}^*_{n_{t}t}\} \subset \mathcal{D}$, where $n_{t}$ represents the number of detected individuals. Here, $\mathcal{D} \subset \mathbb{R}^2$ denotes the study area encompassing all sampled habitats. Each automated detection corresponds to the centroid of the associated bounding box predicted by YOLOv11m. As shown in Table \ref{tab:detection_performance}, the automated detection system exhibits high precision but moderate recall,  resulting in systematic undercounting relative to manual annotation ($n_{t} < m_{t} \ \forall t$ ). This conservative choice reflects the statistical principle that false negatives (Type II errors) can be corrected through modelling, whereas false positives (Type I errors) would introduce spurious spatial patterns that cannot be readily distinguished from true ecological signals.


To correct for this bias and recover the latent spatial distribution of sea cucumbers, we conceptualise the detected pattern $\mathbf{U}^*_t$ as a thinned realisation of the underlying ecological process for each campaign $t$. Specifically, each true individual located at $\mathbf{s} \in \mathcal{D}$ is independently detected with probability $p_t(\mathbf{s}) \in [0,1]$, producing an observed thinned process \citep{dorazio2014accounting}. Under this framework, $\mathbf{U}^*_t$ is modelled as a realisation of a thinned Log-Gaussian Cox Process (LGCP) \citep{moller1998log}:
\begin{equation}
\label{eq:intensity}
\begin{aligned}
    \mathbf{U}^*_t &\mid \lambda^*_t(\mathbf{s}) \sim \mathrm{PP}(\lambda_t^*(\mathbf{s})) \\
    \lambda_{t}^*(\mathbf{s}) &= \lambda_{t}^{\text{pot}}(\mathbf{s}) \, p_{t}(\mathbf{s}) \\
    \log(\lambda_{t}^{\text{pot}}(\mathbf{s})) &= \mu_{t} + \mathbf{x}(\mathbf{s})^\top \boldsymbol{\beta} + w_{l}(\mathbf{s}) \\
    w_{l}(\mathbf{s}) &\sim \mathrm{GP}\!\left(0, C(\cdot; \sigma^2_{l}, \rho_{l})\right), \quad l = 1, 2
\end{aligned}
\end{equation}
Here, $\lambda_t^{\text{pot}}(\mathbf{s})$ denotes the potential intensity, representing the latent ecological distribution of sea cucumbers in the absence of detection bias \citep{chakraborty2011point, warton2010poisson}. The parameter $\mu_{t}$ denotes a campaign-specific intercept, $\mathbf{x}(\mathbf{s})$ is a vector of spatial covariates, and $w_{l}(\mathbf{s})$ is a Gaussian Process. The index $l$ equals $1$ for the first three campaigns and $2$ for the remaining ones. This grouping was chosen because the number of observed points in the latter two campaigns is substantially lower than in the first three. The observed intensity $\lambda_t^*(\mathbf{s})$ thus represents a filtered version of this potential intensity, modulated by the spatially varying detection probability $p_t(\mathbf{s})$. Several formulations of $p_t(\mathbf{s})$ are possible, including detection functions inspired by distance sampling theory \citep{martino2021integration, yuan2017point}. In this study, we adopt a half-normal detection function. 
\begin{align}
    p_{t}(\mathbf{s}) = \exp \left( - \frac{ (z_{t}(\mathbf{s}))^2}{2\tau^2} \right),
\end{align}
where $z_t(\mathbf{s})$ is a covariate affecting detectability and $\tau$ is a scale parameter controlling the rate of detection decay. This formulation assumes maximum detection when the covariate equals zero, with detection probability decreasing as the covariate deviates from this reference value. A key advantage of this framework is its flexibility in accommodating multiple sources of detection heterogeneity. The detection probability can be extended by defining $p_t(\mathbf{s})$ as a product of independent half-normal components:
\begin{align}
    p_{t}(\mathbf{s}) = \prod_{k = 1}^{K}\exp \left( - \frac{ (z_{t, k}(\mathbf{s}))^2}{2\tau_{k}^2} \right),
\end{align}
where $K$ represents the number of covariates employed. Critically, the detection mechanism functions as a spatial filter that preserves the underlying ecological point pattern structure while selectively revealing individuals as a function of covariates.


\subsection{Model Estimation}

Model estimation was performed using the \href{https://www.r-inla.org/home}{R-INLA} package \citep{rue2009inla}, through the \emph{inlabru} interface \citep{bachl2019inlabru}. INLA provides a computationally efficient framework for approximate Bayesian inference in Latent Gaussian Models, characterised by Gaussian latent fields governed by a limited set of hyperparameters and non-Gaussian likelihoods. The approach relies on deterministic Laplace approximations to estimate posterior marginals, offering substantial computational advantages over traditional Markov Chain Monte Carlo (MCMC) methods.

Weakly informative Gaussian priors were assigned to the intercept parameters $\mu_t$ and regression coefficients $\boldsymbol{\beta}$, both centred at zero with precision $0.001$. For the spatial Gaussian processes, penalised complexity (PC) priors \citep{simpson2017penalising} were specified for the marginal standard deviation $\sigma_{l}$ and range $\rho_{l}$ parameters. Specifically, we set $\mathbb{P}(\rho_{l} < 50) = 0.5$ and $\mathbb{P}(\sigma_{l} > 0.5) = 0.01$ for both spatial components, reflecting the spatial scale and variability expected within the study domain. Finally, the scale parameters of the detection functions were modeled through a non-informative Gaussian prior to mitigate numerical instabilities observed during estimation.

\subsection{Model Comparison}\label{mod_comp}

Model selection in a Bayesian context requires evaluation criteria that appropriately balance predictive accuracy and model complexity. Although traditional measures such as the Deviance Information Criterion (DIC) are widely used, they exhibit well-documented limitations, particularly for hierarchical or spatially structured models \citep{gelfand2018bayesian, leininger2017bayesian}. We therefore adopted a predictive evaluation framework grounded in the theory of innovation processes for spatial point patterns \citep{baddeley2005residual}, complemented by proper scoring rules such as the Continuous Ranked Probability Score (CRPS) \citep{matheson1976scoring}. To account for spatial heterogeneity and temporal variation, the study region was partitioned into bounded subdomains $B_{1}, \ldots, B_{G}$. This spatial blocking enables localised residual analysis, providing more detailed model diagnostics across heterogeneous habitats. For each campaign $t$, let $\mathbf{O}_{t} \mid \lambda_{t}$ denote a non-homogeneous Poisson process with intensity $\lambda_t(\mathbf{o})$ defined over the spatial domain $\mathcal{D}$. A realization $\mathbf{O}_t$ can be partitioned into $\mathbf{o}_t^1, \ldots, \mathbf{o}_t^{G}$, corresponding to observations within each subregion. Given an innovation function $h_{t}(\cdot)$, the innovation process in the $g$-th subset at time $t$ is defined as
\begin{align}
R_{h}^{t}(B_{g}) = \sum_{\mathbf{o} \in \mathbf{o}_{t}^{g}} h_{t}(\mathbf{o}) - \int_{B_{g}} h(\mathbf{o})\,\lambda_{t}(\mathbf{o})\,d\mathbf{o}.
\end{align}
Setting $h_{t}(\mathbf{o}) = 1$ yields the raw residuals:
\begin{equation}
R_{\text{raw}}^{t}(B_{g}) = N(B_{g}) - \int_{B_{g}}  \lambda(\mathbf{o})\,d\mathbf{o},
\end{equation}
where $N(B_{g})$ denotes the observed number of points within $B_{g}$.  
Raw residuals measure the difference between observed and expected point counts but may be dominated by regions of high intensity, thereby masking potential model deficiencies in sparser areas.  To mitigate this imbalance, we computed Pearson residuals by setting $h_{t}(\mathbf{o}) = 1 / \sqrt{\lambda_{t}(\mathbf{o})}$:
\begin{align}
R_{\text{pearson}}^t(B_{g}) = \sum_{\mathbf{u} \in \mathbf{u}_{t}^{g}} \frac{1}{\sqrt{\lambda_{t}(\mathbf{u})}} - \int_{B_g} \sqrt{\lambda_{t}(\mathbf{u})}\,d\mathbf{u}.
\end{align}
This normalisation stabilises the variance by weighting each observation inversely to its expected intensity, reducing heteroscedasticity and allowing for balanced residual diagnostics across regions of varying density. Integrals of the intensity function were approximated using Monte Carlo quadrature \citep{berman1992approximating}. Let $\{c_i \in \mathcal{D}\}_{i=1}^N$ denote quadrature nodes and $\{w_i\}_{i=1}^N$ their associated weights. The integral of the intensity over $\mathcal{D}$ was then approximated as
\begin{align}
\int_{\mathcal{D}} \lambda_t(\mathbf{o}) \, do \approx \sum_{i=1}^N w_i \, \lambda_t(c_i),
\end{align}
and this computation was repeated across $M$ posterior samples to characterise the uncertainty in residual-based diagnostics. To complement residual analysis, we employed the CRPS as a proper scoring rule for evaluating predictive distributions. Under correct model specification, residuals should have a zero expectation \citep{baddeley2005residual}. Denoting by $F_{R_{\text{raw},t}}^{g}$ and $F_{R_{\text{pearson},t}}^{g}$ the predictive distributions of raw and Pearson residuals, respectively, the CRPS is defined as
\begin{align}
    \text{CRPS}(F_{R_{\text{raw},t}}^{g}, 0) &= \int_{\mathbb{R}}  (F_{R_{\text{raw},t}}^{g}(x) - \mathbf{1}\{ x \geq 0\})^2 \, dx,\\
    \text{CRPS}(F_{R_{\text{pearson},t}}^{g}, 0) &= \int_{\mathbb{R}}  (F_{R_{\text{pearson},t}}^{g}(x) - \mathbf{1}\{ x \geq 0\})^2 \, dx.
\end{align}

The CRPS offers distinct advantages over information criteria such as DIC, as it assesses the entire predictive distribution rather than a single summary statistic. This makes it particularly appropriate for spatial point process models with hierarchical structures, where predictive calibration and uncertainty quantification are of primary interest.

\section{Results}\label{sec3}

Because false negatives cannot be directly observed, their treatment requires explicit assumptions regarding their dependence on observable covariates. We hypothesise that the probability of false negatives is not uniformly distributed across space but varies systematically according to three factors. First, we expect false negatives to be more frequent in areas of high local density, where individuals may overlap or occlude one another. 
To capture this effect, we define the local frequency ($f_{r}$) as the number of predicted points within a circle of radius $r$ centred
on each detection. The choice of spatial scale $r$ for computing local frequency requires justification based on the spatial structure of our data. To select an appropriate radius, we examined the radius that guarantees a good percentage of detected points have at least one neighbour, finding that a good threshold is approximately $ r = 1$, which guarantees approximately $85\%$ coverage. 
Second, we anticipate that the detection probability is related to the confidence score ($CS$) assigned by the neural network detector, which ranges from $0$ to $1$. In fact, individuals with lower confidence scores are more likely to represent marginal detections in regions of poor algorithm performance. Third, we expect a relationship with object size, approximated by the diagonal length ($DL$) of each bounding box, also ranging from $0$ to $1$. Smaller individuals are inherently more difficult to detect.
To incorporate these covariates into the half-normal detection function, we applied transformations to ensure each acts appropriately as a detectability modifier. Specifically, we used the complement to one for the confidence score and the bounding box diagonal and the reciprocal for the local frequency. Preliminary exploratory analysis [qui metterei altro articolo] suggested that detection errors were approximately evenly distributed across habitat categories. 


\begin{table}[t]
\caption{Comparison of Methods Across Five Campaigns using raw residuals (upper table) and Pearson residuals. Model $1$ is the model fitted on the observed values, while Model $2$ is the model fitted on the detected locations without thinning.}
\label{tab:residuals}
\renewcommand{\arraystretch}{1.2}
\centering
\begin{tabular}{l
                S[table-format=1.3]
                S[table-format=1.3]
                S[table-format=1.3]
                S[table-format=1.3]
                S[table-format=1.3]
                S[table-format=1.3]}
\toprule
\textbf{Method} & \textbf{$t=1$} & \textbf{$t=2$} & \textbf{$t=3$} & \textbf{$t=4$} & \textbf{$t=5$} & \textbf{Average} \\
\midrule
{Model 1}      & 4.356 & 4.441 & 4.536 &3.369 & 3.250 & 3.990\\
Model 2      & 5.867 & 6.929 & 5.912 & 6.246 & 8.475 & 6.694\\
Model 3 diag & 5.558 & 6.444 & 5.551 & 5.998 & 7.948 & 6.303 \\
Model 3 freq & 5.486 & 6.383 & 5.480 & 5.937 & 7.867 & 6.230 \\
Model 3 conf &5.484 & 6.376 & 5.505 & 5.914 & 7.791 & 6.214\\
Model 3 conf $\times$ freq &  5.420 & 6.313 & 5.438 &  5.864 & 7.696 & \textbf{6.150}\\
\bottomrule
\end{tabular}

\vspace{1em}

\begin{tabular}{l
                S[table-format=1.3]
                S[table-format=1.3]
                S[table-format=1.3]
                S[table-format=1.3]
                S[table-format=1.3]
                S[table-format=1.3]}
\toprule
\textbf{Method} & \textbf{$t=1$} & \textbf{$t=2$} & \textbf{$t=3$} & \textbf{$t=4$} & \textbf{$t=5$} & \textbf{Average} \\
\midrule
Model 1      & 1.615 & 1.905 & 1.839 & 1.152 & 0.998& 1.500\\
Model 2      & 1.964 & 2.685 & 2.101 & 1.803 & 1.911& 2.092 \\
Model 3 diag & 1.887 & 2.540 & 2.021 & 1.749 & 1.837& 2.010 \\
Model 3 freq & 1.865 & 2.515 & 1.997 & 1.740 & 1.825& 1.999 \\
Model 3 conf &1.864 & 2.516 & 2.004 & 1.734 & 1.816 & 1.992\\
Model 3 conf $\times$ freq& 1.852 & 2.493 & 1.987 &  1.724 & 1.802 & \textbf{1.979}\\
\bottomrule
\end{tabular}
\end{table}

Across all models, the potential intensity function was specified using the same set of covariates, namely the percentage cover of P. \textit{Oceanica}, the slope of the seafloor and a Gaussian process component to capture residual spatial variation. We fitted a model based on the observed locations, a second model using the detected locations, and a series of models incorporating the thinning mechanism applied to the detected subset. The latter included models using $f_{r}$, $CS$, and $DL$ individually, as well as their combinations. Residuals were computed on a localised $18 \times 18$ grid superimposed on the study area, following the procedure described in Subsection~\ref{mod_comp}. Table \ref{tab:residuals} shows the average value of the CRPS for the four best thinning models, together with those for the two baseline models, using both Pearson and raw residuals. Several patterns emerge from this comparison. First, Model 2 (fitted to unadjusted detections) shows substantially worse performance than Model 1 (observed data), with average Pearson residuals of $6.694$ compared to $3.990$ and raw residual of $2.092$ compared to $1.500$. This confirms that ignoring false negatives in the ground truth introduces considerable bias in spatial intensity estimation. Second, all thinning models systematically improve performance relative to Model 2. The improvement is consistent across all five temporal campaigns, suggesting that the thinning mechanism successfully corrects for spatially structured false negatives. Third, among the thinning specifications tested, the model incorporating both local frequency and confidence score through a product of half-normal functions (Model 3 conf $\times$ freq) achieves the best performance. This suggests that false negatives are most strongly associated with local crowding and marginal detections, as indicated by low confidence scores. However, none of the thinning models fully recovers the performance of Model $1$. Nevertheless, the substantial improvement over Model $2$ demonstrates that the thinning framework provides a practical method for reducing bias when ground truth data contain false negatives.

\begin{figure}[t]
    \centering
     \subfloat[]{\includegraphics[scale = 0.26]{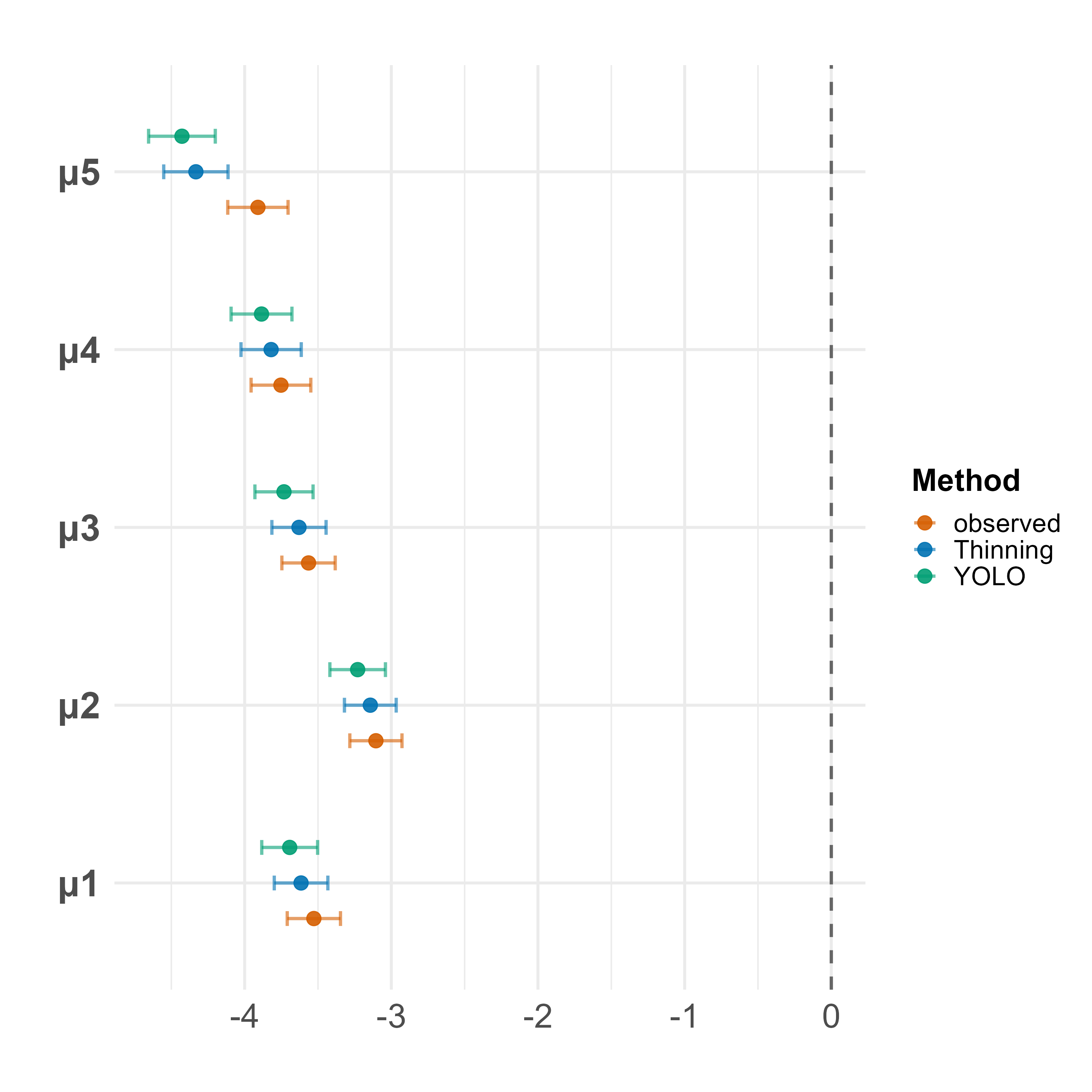}}
     \subfloat[]{\includegraphics[scale = 0.26]{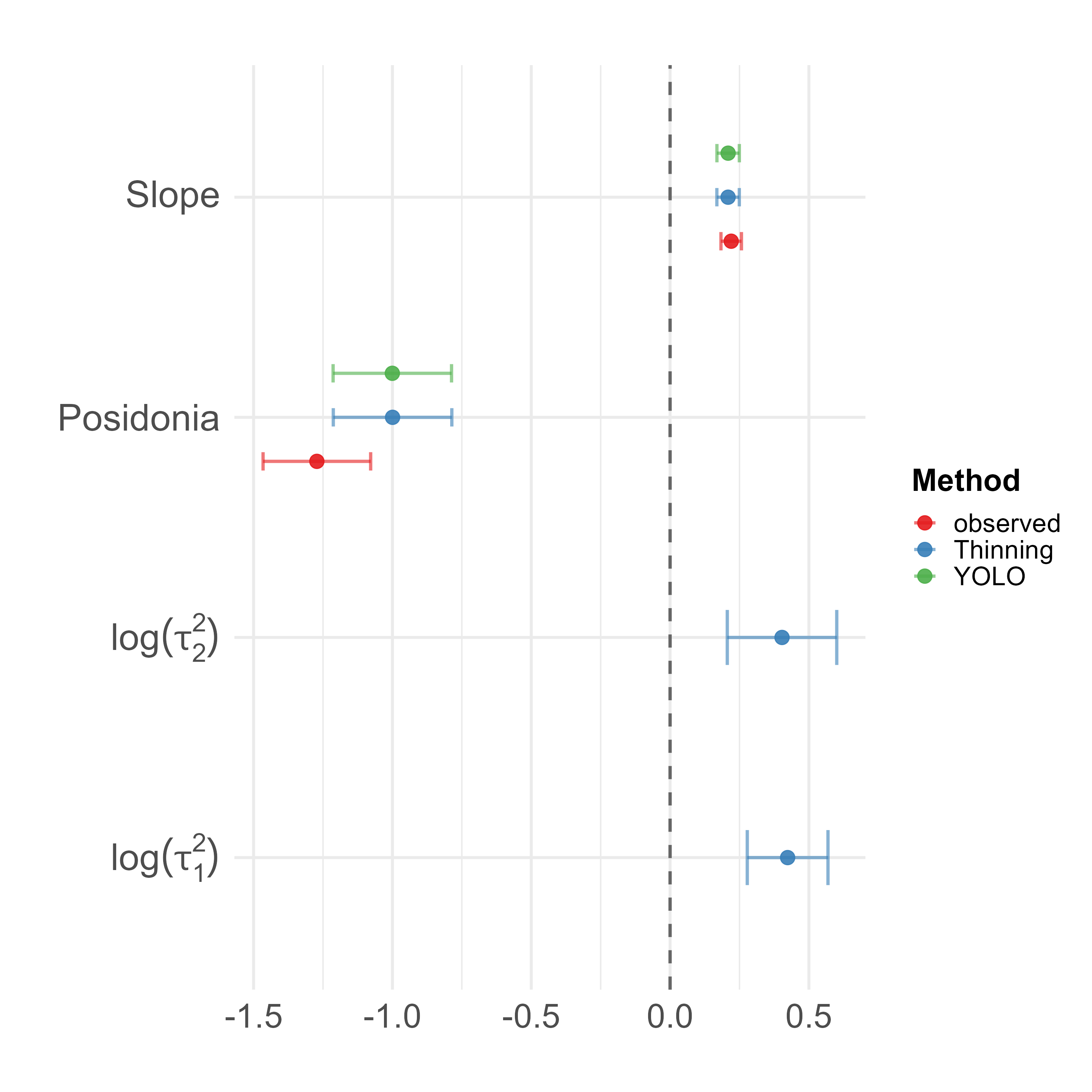}}
     \caption{Posterior distributions of the parameters representing mean and $95\%$ credible intervals involved in the model fitted on the observed value, the model fitted on the detection and the best model with thinning. }
     \label{fig:param}
\end{figure}

To assess whether the thinning approach recovers the underlying ecological relationships, we compared the posterior distributions of the intensity function parameters across Model $1$, Model $2$, and the best thinning model. Figure \ref{fig:param} displays the posterior means and $95\%$ credible intervals for the seasonal intercepts, the environmental covariates and the scale parameters of the half-normal. For the seasonal intercepts, Model 3 shows closer agreement with Model 1 than does Model 2. This suggests that accounting for false negatives through thinning partially recovers the true temporal variation in population intensity. For the environmental covariates the posterior distributions are similar across Model 2 and Model 3. This indicates that while false negatives bias the absolute intensity estimates (captured by the intercepts), they have less impact on the relative effects of environmental covariates. We can conclude that the thinning correction altered the magnitude of the estimated intensities, without modifying the spatial allocation.

The posterior distributions of the thinning function scale parameters reveal moderate thinning, and the similar magnitudes of both scale parameters suggest that local crowding and detection confidence contribute approximately equally to the spatial pattern of false negatives.


\section{Discussion}\label{sec4}

Our methodological framework is fundamentally grounded on the assumption that we are working with a sample of the true population. This conservative stance underpins our decision to avoid measurement error models \citep{chakraborty2010analyzing}, as our data do not exhibit the spatial shifts or misalignments that would justify the additional complexity of such models. By adopting this perspective, we prioritise a transparent and interpretable modelling approach while acknowledging the limitations inherent in working with partial observations.

Central to our approach is the specification of a thinning function designed to account for detection errors in the observed point pattern. While our chosen function provides a tractable mechanism for correcting false positive detections, several alternative specifications were considered. One such alternative involves the integration of multiple covariates through a logit link, resulting in a “pogit” formulation \citep{dorazio2014accounting, winkelmann1993poisson}. To maintain identifiability under this formulation, it is necessary that the covariates $x(\mathbf{s})$ and $z(\mathbf{s})$ remain uncorrelated \citep{fithian2012finite}. However, preliminary exploration revealed that this approach produced excessive thinning, failing to correct for false negative rates adequately, and thereby compromising model performance. Another idea considered, though ultimately not implemented, involves introducing an additional Gaussian process within the thinning mechanism, taking values approximately equal to zero in regions where no counts were observed. Conceptually, this would allow the model to represent areas with minimal or absent detection explicitly. The main impediment to implementing this strategy in our study is the lack of ancillary information to guide the latent field, as was available in \cite{jones2018spatiotemporal}. Without such information, the additional process would be poorly constrained, increasing the risk of identifiability issues and overfitting. A further alternative involves non-parametric thinning mechanisms, which have thus far been applied primarily in the context of discrete count data \citep{arima2024bayesian}. These mechanisms are appealing because they allow heterogeneous thinning patterns driven by multiple covariates, potentially capturing complex detection behaviours that parametric models cannot. While adaptation to thinned LGCPs is theoretically feasible, methodological development is still required to implement such models in continuous spatial point process contexts. Consequently, this remains a promising but largely conceptual avenue for future research.

Several intrinsic limitations arise from our current model specification. The effectiveness of the thinning mechanism is inherently linked to the underlying object detector’s performance. Our framework prioritises minimising false positives, which renders it highly sensitive to false negatives. Once false negative rates exceed a certain threshold, the model’s capacity to correct for detection errors diminishes substantially. Moreover, while our approach effectively accounts for directly observable false positives, unobserved missed detections represent a more formidable challenge. This limitation suggests that latent variable formulations, akin to those used in degraded point pattern analysis \citep{chakraborty2011point}, could provide a pathway to incorporate unobserved events into the model framework.

Our methodology represents a departure from traditional approaches in the field, which often combine detection and spatial processes in a single modeling stage. By explicitly separating detection uncertainty from spatial modeling and employing the thinning mechanism to quantify this uncertainty, we provide a clearer delineation between observational errors and underlying spatial structure. Nonetheless, our framework does not explicitly propagate detection uncertainty using formal error propagation techniques. Future work could leverage conformal prediction \citep{vovk2005algorithmic, deliu2025interplay} or related methods to define more informative priors on detection function parameters, improve uncertainty quantification, and enhance the theoretical foundations of the approach. Such developments would strengthen methodological rigor and provide practitioners with more reliable estimates of uncertainty, advancing the capacity to model imperfect detection in spatial point processes.

Finally, we note a related consideration that, while ancillary to the present work, merits acknowledgment. Model selection in this context is inherently problematic. Standard cross-validation procedures are fundamentally unsuited to these applications because we do not observe the true underlying point pattern. The observed data themselves are products of the detection process we are attempting to model, creating a circularity: evaluating predictive performance against these observations conflates genuine model fit with artifacts of the thinning mechanism itself. This is the rationale behind fitting multiple candidate models and evaluating their residuals. This residual-based approach assesses the quality of the fitted spatial field without requiring holdout predictions on the thinned observations, thereby avoiding the logical inconsistency inherent in standard cross-validation for thinned point processes. While not a complete solution to the model selection challenge, this strategy provides a pragmatic means of comparing competing specifications. Alternative selection criteria that more formally account for the dual structure of intensity and detection processes warrant deeper investigation in future methodological development.


From an ecological perspective, correctly identifying the spatial distribution of benthic populations such as sea cucumbers and sea urchins is essential for understanding population resilience and informing management strategies.
Previous work based on in-field visual census data \citep{addis2012spatial} has shown that sea urchin population often display strong small-scale autocorrelation and patchy distribution patterns. Consequently, a spatial model's capacity to estimate detection probability while concurrently recovering the actual underlying spatial intensity and assessing its relationship with environmental covariates can serve as a
significant resource for monitoring benthic communities. 
The increasing reliance on automated image-based surveys and underwater photogrammetry amplifies this challenge. Despite this, recent efforts to map the densities of sea urchins
\citep{piazza2019underwater, sastraantara2024mapping},
macroalgae canopy, and the coverage of other benthic species \citep{spyksma2022diver} have alleviated
the constraints of monitoring budgets and timelines by delivering cost-effective, high-quality data
across significantly expanded temporal and spatial scales. Nevertheless, several challenges persist in
the rapid assessment of spatial patterns. Indeed, the integration of underwater photogrammetry
with unmanned platforms, such as unmanned aerial vehicles (UAV), for shallow water applications
\citep{sugara2025detection} frequently results in extensive imagery datasets that must be analyzed using
deep learning-based detectors, which may inadvertently introduce bias in detection.




In summary, our approach balances tractability, interpretability and methodological rigor, while acknowledging inherent limitations. The obtained results demonstrate that false negatives in ground truth data can substantially bias spatial intensity estimates when using automated detection systems. The thinning framework provides a principled approach to correcting for this bias, particularly when false negatives follow predictable spatial patterns related to local density and detection confidence. While the correction does not fully eliminate the bias introduced by imperfect ground truth, it represents a considerable improvement over naive approaches that ignore detection errors in training data. Future applications should carefully consider which covariates best capture the structure of false negatives in their specific study systems. We have carefully considered alternative specifications and potential extensions, which may guide future research, but our current formulation represents a pragmatic compromise tailored to the available data and the specific characteristics of the object detection system.

\backmatter

\bmhead{Acknowledgements}
The authors would like to thank Daniele Poggio for some stimulating conversation on the topic.

\bibliography{sn-bibliography}

\end{document}